\documentclass[]{aastex631}
\usepackage{bm}

\begin{document}

\title{Inertial-range Turbulence Anisotropy of the Young Solar Wind from Different Source Regions}

\author[0009-0005-3941-1514]{Wenshuai Cheng}
\affiliation{State Key Laboratory of Solar Activity and Space Weather, National Space Science Center, Chinese Academy of \\
Sciences, Beijing, China; \href{mailto:mxiong@swl.ac.cn}{mxiong@swl.ac.cn}, \href{chengwenshuai22@mails.ucas.ac.cn}{chengwenshuai22@mails.ucas.ac.cn}}

\affiliation{University of Chinese Academy of Sciences, Beijing, China}

\author[0000-0001-9427-7366]{Ming Xiong}
\affiliation{State Key Laboratory of Solar Activity and Space Weather, National Space Science Center, Chinese Academy of \\
Sciences, Beijing, China; \href{mailto:mxiong@swl.ac.cn}{mxiong@swl.ac.cn}, \href{chengwenshuai22@mails.ucas.ac.cn}{chengwenshuai22@mails.ucas.ac.cn}}

\affiliation{University of Chinese Academy of Sciences, Beijing, China}

\author[0009-0004-4832-0895]{Yiming Jiao}
\affiliation{State Key Laboratory of Solar Activity and Space Weather, National Space Science Center, Chinese Academy of \\
Sciences, Beijing, China; \href{mailto:mxiong@swl.ac.cn}{mxiong@swl.ac.cn}, \href{chengwenshuai22@mails.ucas.ac.cn}{chengwenshuai22@mails.ucas.ac.cn}}

\affiliation{University of Chinese Academy of Sciences, Beijing, China}

\author[0000-0002-8234-6480]{Hao Ran}
\affiliation{Mullard Space Science Laboratory, University College London, Dorking, RH5 6NT, United Kingdom}

\author[0000-0003-4716-2958]{Liping Yang}
\affiliation{State Key Laboratory of Solar Activity and Space Weather, National Space Science Center, Chinese Academy of \\
Sciences, Beijing, China; \href{mailto:mxiong@swl.ac.cn}{mxiong@swl.ac.cn}, \href{chengwenshuai22@mails.ucas.ac.cn}{chengwenshuai22@mails.ucas.ac.cn}}

\author[0000-0001-8188-9013]{Huidong Hu}
\affiliation{State Key Laboratory of Solar Activity and Space Weather, National Space Science Center, Chinese Academy of \\
Sciences, Beijing, China; \href{mailto:mxiong@swl.ac.cn}{mxiong@swl.ac.cn}, \href{chengwenshuai22@mails.ucas.ac.cn}{chengwenshuai22@mails.ucas.ac.cn}}

\author[0000-0001-5205-1713]{Rui Wang}
\affiliation{State Key Laboratory of Solar Activity and Space Weather, National Space Science Center, Chinese Academy of \\
Sciences, Beijing, China; \href{mailto:mxiong@swl.ac.cn}{mxiong@swl.ac.cn}, \href{chengwenshuai22@mails.ucas.ac.cn}{chengwenshuai22@mails.ucas.ac.cn}}

\affiliation{University of Chinese Academy of Sciences, Beijing, China}

\begin{abstract}
We investigate the wavevector and variance anisotropies in the inertial range of the young solar wind observed by the Parker Solar Probe (PSP). Using the first 19 encounters of PSP measurements, we identify the young solar wind from different source regions: coronal hole (CH) interiors, streamers, and low Mach-number boundary layers (LMBLs), i.e., the peripheral region inside CHs. We assess the wavevector anisotropy with the 2D and slab turbulence model for the CH wind and the streamer wind, and the nearly incompressible (NI) MHD turbulence model for the LMBL wind where Taylor’s hypothesis becomes questionable. Unlike the $\sim80\%$ 2D contribution typically reported at 1 au, our results show that only $26\%$ of the inertial range energy is associated with 2D fluctuations in the CH wind, and this fraction increases to $45\%$ in the streamer wind. As a representation of the LMBL wind, similarly, the oblique sub-Alfvénic intervals and the near-subsonic intervals are characterized by the dominance of slab fluctuations. All the results suggest that slab fluctuations are more abundant in the young solar wind below 0.3 au than at 1 au. Furthermore, we find a dependence of the variance anisotropy in the inertial range on proton plasma beta $\beta_p$. The variance anisotropy is the strongest in the LMBL wind with the lowest $\beta_p$, and the weakest in the streamer wind with the highest $\beta_p$. This contrast can be interpreted as the remnant of fluctuations from the coronal sources.
\end{abstract}

\section{Introduction} \label{sec:Intro} 
An important characteristic of turbulence in the solar wind is the wavevector anisotropy introduced by a mean magnetic field \citep{2012SSRv..172..325H, 2015RSPTA.37340152O}. The wavevector anisotropy refers to unequal distribution of power relative to the direction of the wavevector, which is associated with the anisotropic transfer of energy in the MHD turbulent cascade \citep[e.g.,][]{1981PhFl...24..825M, 1982PhST....2...83M, 1983JPlPh..29..525S, 1995ApJ...438..763G}. Based on the two-dimensional correlation function of solar wind fluctuations at 1 au, \cite{1990JGR....9520673M} proposed that the magnetic energy is preferentially distributed between field‐aligned and perpendicular wavevectors. This naturally led to a two-component model comprising both slab and 2D fluctuations \citep{1992JGR....9717189Z, 1993PhFlA...5..257Z}. In order to further investigate the wavevector anisotropy, \cite{1996JGR...101.2511B} developed a method to quantify the power fractions of the slab and 2D components via Taylor’s hypothesis \citep{1935RSPSA.151..421T}. This is the basis of our subsequent analysis. Their results suggested that the 2D component was energetically dominant at frequencies $10^{-4}-10^{-2}$ Hz with $\sim 80 \%$ of the energy, determined from Helios observations at 1 au. In light of the method provided by \cite{1996JGR...101.2511B}, several studies also found an energetic predominance for the 2D component in the high frequency end of the inertial range from $0.3-1$ au \citep{1998JGR...103.4775L, 2008JGRA..113.1106H, 2010JGRA..115.7105M}. Notably, there is a distinction in the distribution of energy across wavevectors between the fast wind and the slow wind at 1 au, which can be interpreted as a remnant of the solar source \citep{2005ApJ...635L.181D, 2008JGRA..113.1106H}. However, the original features of slab and 2D fluctuations in the young solar wind from different source regions are not well known.

The launch of the Parker Solar Probe \citep[PSP;][]{2016SSRv..204....7F} mission in 2018 makes it possible to investigate the wavevector anisotropy of the young solar wind within 0.3 au from the Sun, where the original features of the fluctuations are most clearly detectable. In the new regime being explored by PSP, the early results on the wavevector anisotropy in the inertial range describe the nascent solar wind evolution, although there is no classification for fluctuations according to the origin of the solar wind due to insufficient data. Applying the method of \cite{1996JGR...101.2511B} to the first five encounter measurements of PSP, \cite{2021ApJ...923..193B} suggested that the relative power in 2D fluctuations is smaller closer to the Sun. Similar results were obtained by \cite{2022ApJ...924L...5Z}. With a broader dataset now available, extensive wind intervals from different source regions can be identified from the PSP measurements from encounters 1 to 19. These observations provide an opportunity to analyze the wavevector anisotropy for each type of the young solar wind according to its source. Typically, the fast wind is believed to come from inside coronal holes \citep[CHs; e.g.,][]{1976SoPh...46..303N, 1977RvGSP..15..257Z}, whereas the slow wind is thought to arise from closed streamers \citep[e.g.,][]{1997ApJ...484..472S, 2009ApJ...691.1936C} and coronal hole boundaries \citep[e.g.,][]{2015ApJ...805...84D, 2019MNRAS.483.4665D}. A special structure originating from the boundary of a coronal hole with rapidly diverging field lines is proposed by \cite{2023ApJ...944..116L}, termed as a low Mach-number boundary layer (LMBL). The LMBL wind is a unique component in the pristine solar wind observed by PSP, which can explain the observations of the sub-Alfvénic wind and the near-subsonic wind \citep{2023ApJ...944..116L, 2024ApJ...963...85L, 2024ApJ...963...82R, 2024ApJ...967...58C}. Recent works have shown that the CH wind, the LMBL wind, and the streamer wind constitute three typical components of the young solar wind near the Sun \citep{2024ApJ...975L..41J, 2024ApJ...960...42J}. In this work, we classify the young solar wind into the above three types based on the criteria established by \cite{2024ApJ...975L..41J}, and study the wavevector anisotropy of each type of the young solar wind individually.

Since the LMBL wind is mostly sub-Alfvénic and Taylor’s hypothesis becomes questionable, the method of \cite{1996JGR...101.2511B} may not be applicable for the LMBL wind. \cite{2022ApJ...926L..16Z} provided a method to extend the evaluation of the wavevector anisotropy in the sub-Alfvénic solar wind based on the nearly incompressible (NI) MHD turbulence model. The NI MHD spectral theory focuses on the Elsässer variables $\textit{\textbf{z}}^{\pm}$, including a 2D component and a slab component to interpret the observations \citep{2017ApJ...835..147Z, 2020ApJ...900..115Z}. In the analysis of the sub-Alfvénic solar wind at encounter 8, the 2D component can not contribute significantly to the observed turbulence due to the approximately aligned solar wind flow and mean magnetic field, making it challenging to assess the 2D contribution accurately \citep{2022ApJ...926L..16Z}. However, for the wind interval with a oblique sampling angle, both the 2D and slab components are important, and then the full anisotropy can be determined in the NI MHD turbulence model. The highly oblique nature was observed in the sub-Alfvénic interval at encounter 10, where the $\textit{\textbf{z}}^{+}$ spectrum is dominated by the slab component \citep{2022ApJ...934L..36Z}. As PSP dives deeper into the corona, the wavevector anisotropy of the newly observed sub-Alfvénic intervals with oblique sampling angles remains to be assessed, serving as a representation of the young solar wind originating from LMBLs. Moreover, the near-subsonic interval at encounter 15 is also characterized by the dominance of slab fluctuations \citep{2024ApJ...972..129E, 2025ApJ...978L..34Z}. Given the common source region from LMBLs, it is crucial to examine whether all the near-subsonic intervals exhibit the wavevector anisotropy similar to that of the sub-Alfvénic intervals.

Apart from the wavevector anisotropy, another manifestation that the magnetic field affects the solar wind turbulence is the variance anisotropy, first introduced by \cite{1971JGR....76.3534B}. The variance anisotropy is determined by the relative magnitudes of the energy in the magnetic field fluctuations perpendicular and parallel to the mean magnetic field, which is often associated with magnetic compressibility \citep[e.g.,][]{1998JGR...103.4775L, 2008JGRA..113.1106H, 2010JGRA..115.7105M, 2020ApJ...900...93P}. The variance anisotropy appears to be modulated by the proton plasma beta $\beta_p$ \citep{1996JGR...101.7619M}. \cite{2001JGR...10618625S} reported a near-Earth sub-Alfvénic interval with stronger variance anisotropy in the inertial range and a lower $\beta_p$ than the neighboring super-Alfvénic intervals. Similar results were obtained in the sub-Alfvénic intervals observed by PSP \citep{2022ApJ...926L...1B}. In a subsequent statistical analysis spanning a broad range of solar wind conditions, \cite{2006JGRA..111.9111S} found that the variance anisotropy in the inertial range is inversely correlated with $\beta_p$ at 1 au. This correlation follows a power-law dependence $\sim \beta_p^{-0.7}$ and holds down to 0.3 au \citep{2010JGRA..115.7105M}. The relationship between the variance anisotropy and the $\beta_p$ in the young solar wind within 0.3 au has yet to be investigated. 

In this Letter, we focus on the wavevector and variance anisotropies of the young solar wind and examine how different source regions shape the original features of the anisotropic fluctuations. In Section \ref{sec:DAM}, we introduce the dataset and the methods used to analyze the wavevector anisotropy in the young solar wind from different source regions in detail. In Section \ref{sec:OAR}, we give an overview as an example about dividing the young solar wind into three source regions, including CH interiors, streamers, and CH boundaries. Then we quantify the wavevector anisotropy for the CH wind and the streamer wind using the \cite{1996JGR...101.2511B} analysis. For the LMBL wind, where Taylor’s hypothesis is no longer valid, we apply the method developed by \cite{2022ApJ...926L..16Z} to compare the contributions from the slab and 2D components. We also find a dependence of the variance anisotropy on the $\beta_p$ in the young solar wind within 0.3 au. The conclusions are summarized in Section \ref{sec:Conclusion}.

\section{Data and Methodology} \label{sec:DAM}
\subsection{PSP Data} \label{sec:dataset}
The measurements in this study are provided by both the FIELDS instrument suite \citep{2016SSRv..204...49B} and the SWEAP package \citep{2016SSRv..204..131K} aboard PSP. The proton data are from the ion electrostatic analyzer \citep[SPAN-I;][]{2022ApJ...938..138L} and the Faraday cup \citep[the Solar Probe Cup (SPC);][]{2020ApJS..246...43C}. We only use the proton temperature $T_p$ from SPAN-I in the following study. In addition, parameters of the alpha particles are available starting from the fourth encounter. The magnetic field data are obtained from the FIELDS fluxgate magnetometers. The electron density and core temperature are obtained from the quasi-thermal noise (QTN) measurements \citep{2020ApJS..246...44M}. We use the QTN density to represent the plasma density in order to have the most accurate density determination. All the solar wind parameters are set to a cadence of 5 s unless otherwise stated. Our analysis uses the first 19 encounters measurements of PSP to select solar wind data from different source regions. Encounter 11 is not included due to a lack of QTN data. For the identification of the CH wind, the streamer wind and the LMBL wind, we adopt the criteria proposed by \cite{2024ApJ...975L..41J}. 

We compute the power spectra of magnetic fluctuations for each wind type, and the intervals for analysis are chosen to be 3 hr in length spanning several correlation lengths. We first rotate the magnetic field data in RTN coordinates with a cadence of $\sim$ 0.22 s into a mean field coordinate system for each interval \citep[e.g.,][]{2022ApJ...926L..16Z, 2024ApJ...973...26Z}. The orthogonal mean field coordinates ($\hat{\textit{\textbf{x}}}$, $\hat{\textit{\textbf{y}}}$, $\hat{\textit{\textbf{z}}}$) are the normalized ($\langle \textit{\textbf{B}} \rangle$ $\times$ $\langle \bm{V^{sc}} \rangle$ $\times$ $\langle \textit{\textbf{B}} \rangle$, $\langle \textit{\textbf{B}} \rangle$ $\times$ $\langle \bm{V^{sc}} \rangle$, $\langle \textit{\textbf{B}} \rangle$), where $\langle \textit{\textbf{B}} \rangle$ is the mean magnetic field vector and $\langle \bm{V^{sc}} \rangle$ is the mean wind velocity vector in the spacecraft frame. We then calculate the power spectral density (PSD) of the magnetic field components using a standard Fourier transform method. Normally, the range 0.01 Hz to 0.1 Hz in the spacecraft frame is taken to characterize the inertial-range fluctuations. When a power-law form is clearly observed, we fit the above frequency range for each component (i.e., $P_{xx}$, $P_{yy}$, and $P_{zz}$) and the trace of the magnetic spectrum to a power law. The fitted power-law index and the amplitudes of the fluctuations calculated by integrating the PSD over the range $0.01\ \mathrm{Hz} - 0.1\ \mathrm{Hz}$ are used in the following analysis. 
\subsection{Evaluation of the Wavevector Anisotropy} \label{sec:evaluation}
Here, we approximate the anisotropy in the solar wind as the superposition of slab and 2D fluctuations based on the 2D and slab turbulence model \citep[e.g.,][]{1992JGR....9717189Z, 1993PhFlA...5..257Z, 1996JGR...101.2511B, 2017ApJ...835..147Z}. The mean value and standard deviation of the radial Alfvén Mach number are 2.78 ${\pm}$ 1.13 for the CH wind and 4.08 ${\pm}$ 1.97 for the streamer wind, indicating that Taylor’s hypothesis is marginally satisfied. We employ the method proposed by \cite{1996JGR...101.2511B} to investigate the wavevector anisotropy for both the CH wind and the streamer wind:
\begin{equation}
    \frac{P_{yy}}{P_{xx}} = \frac{(1 + q)C_s\cos^{q-1}\theta_{BV} + 2qC_2\sin^{q-1}\theta_{BV}}{(1 + q)C_s\cos^{q-1}\theta_{BV} + 2C_2\sin^{q-1}\theta_{BV}},
    \label{bieber}
\end{equation}
where $q$ is the power-law index of the magnetic trace spectrum, $C_s$ and $C_2$ are the energy in the slab and 2D components, and $\theta_{BV}$ is the angle between the mean magnetic field \bm{$B_{0}$} and mean velocity \bm{$V_{0}^{sc}$} in the spacecraft frame. Equation (\ref{bieber}) shows that more 2D fluctuations are projected along the sampling direction as $\theta_{BV}$ increases, which contributes to the difference between $P_{xx}$ and $P_{yy}$. Then, we can infer the ratio of $C_2$ and $C_s$ by investigating the dependence of the ratio $P_{yy}/P_{xx}$ on the sampling angle $\theta_{BV}$ from Equation (\ref{bieber}).

However, the method of \cite{1996JGR...101.2511B} may not be applicable for the LMBL wind. The mean value and standard deviation of the radial Alfvén Mach number are 0.87 ${\pm}$ 0.50 for the LMBL wind.  In such a sub-Alfvénic regime, the standard Taylor’s hypothesis cannot be used. A modified Taylor's hypothesis can be provided within the context of the NI MHD theory \citep{2022ApJ...926L..16Z}. By considering the Doppler effect of the slab fluctuations, the NI MHD turbulence model can extend the evaluation of the wavevector anisotropy in the sub-Alfvénic wind. Therefore, we adopt the method described in \cite{2022ApJ...926L..16Z} in order to compare the contributions from the slab and 2D components in the LMBL wind. First, we calculate the Elsässer variables defined as $\textit{\textbf{z}}^{\pm} = \delta \textit{\textbf{v}} \mp \mathrm{sign} (\langle B_R \rangle) \delta \textit{\textbf{b}}$, where $\langle B_R \rangle$ is the mean radial magnetic field in RTN coordinates, and $\delta \textit{\textbf{v}}$ and $\delta \textit{\textbf{b}}$ are the fluctuations in the velocity and magnetic field (expressed in Alfvén units). We use the ensemble average to calculate the background mean velocity $V_{0}$ and magnetic field $B_{0}$, which are then subtracted respectively to obtain $\delta \textit{\textbf{v}}$ and $\delta \textit{\textbf{b}}$. The trace spectra of the observed $\textit{\textbf{z}}^{\pm}$ are computed by applying a Fourier transform to each component. The theoretical predictions of the $\textit{\textbf{z}}^{\pm}$ spectra based on the NI MHD turbulence model are given by \cite{2022ApJ...926L..16Z}:
\begin{equation}
    P^{+}(f) = C^{\infty}\frac{2\pi}{|V_0^{sc}\sin\theta_{BV}|}k_{\perp}^{-5/3} \\
    + C^{*+}\frac{2\pi}{|v_{A0} + V_0^{sc}\cos\theta_{BV}|}k_{+}^{-3/2},
    \label{k+}
\end{equation}
\begin{equation}
    P^{-}(f) = C^{\infty}\frac{2\pi}{|V_0^{sc}\sin\theta_{BV}|}k_{\perp}^{-5/3} \\
    + C^{*-}\frac{2\pi}{|v_{A0} - V_0^{sc}\cos\theta_{BV}|}k_{-}^{-3/2} \left( 1 + \left(\frac{k_{-}}{k_{t}}\right)^{1/2} \right)^{1/2},
    \label{k-}
\end{equation}
where $C^{*\pm}$ are the power of the slab component in the forward and backward propagating modes, $C^{\infty}$ is the power of the 2D component (assumed to be equal in both modes), and $k_t$ ($f_t$) is the transition wavenumber (frequency) representing the transition from a regime dominated by nonlinear interactions to Alfvénic interactions. According to Equations (\ref{k+}) and (\ref{k-}), the frequency spectra of $\textit{\textbf{z}}^{\pm}$ can be expressed as the wavenumber spectra of the non-propagating 2D component and the forward (backward) slab component propagating at the Alfvén speed. The wavenumber of the forward (backward) slab fluctuations is calculated as $k_{\pm}=\frac{2\pi f}{|v_{A0}\pm V_0^{sc}\cos\theta_{BV}|}$ considering the Doppler effect, while the wavenumber of the 2D fluctuations is given by $k_{\perp}=\frac{2\pi f}{|V_0^{sc}\sin\theta_{BV}|}$. Here, $v_{A0}$ is the mean Alfvén speed. We then fit the observed $\textit{\textbf{z}}^{\pm}$ spectra based on Equations (\ref{k+}) and (\ref{k-}). From this fitting, we determine the power of the slab components $C^{*\pm}$, the power of the 2D component $C^{\infty}$, and the transition wavenumber (frequency) $k_t$ ($f_t$). Finally, we have a complete set of the description of the wavevector anisotropy for all three wind types.

\section{Overview and Results} \label{sec:OAR}
\begin{figure}[ht]
    \centering
    \includegraphics[scale=0.78]{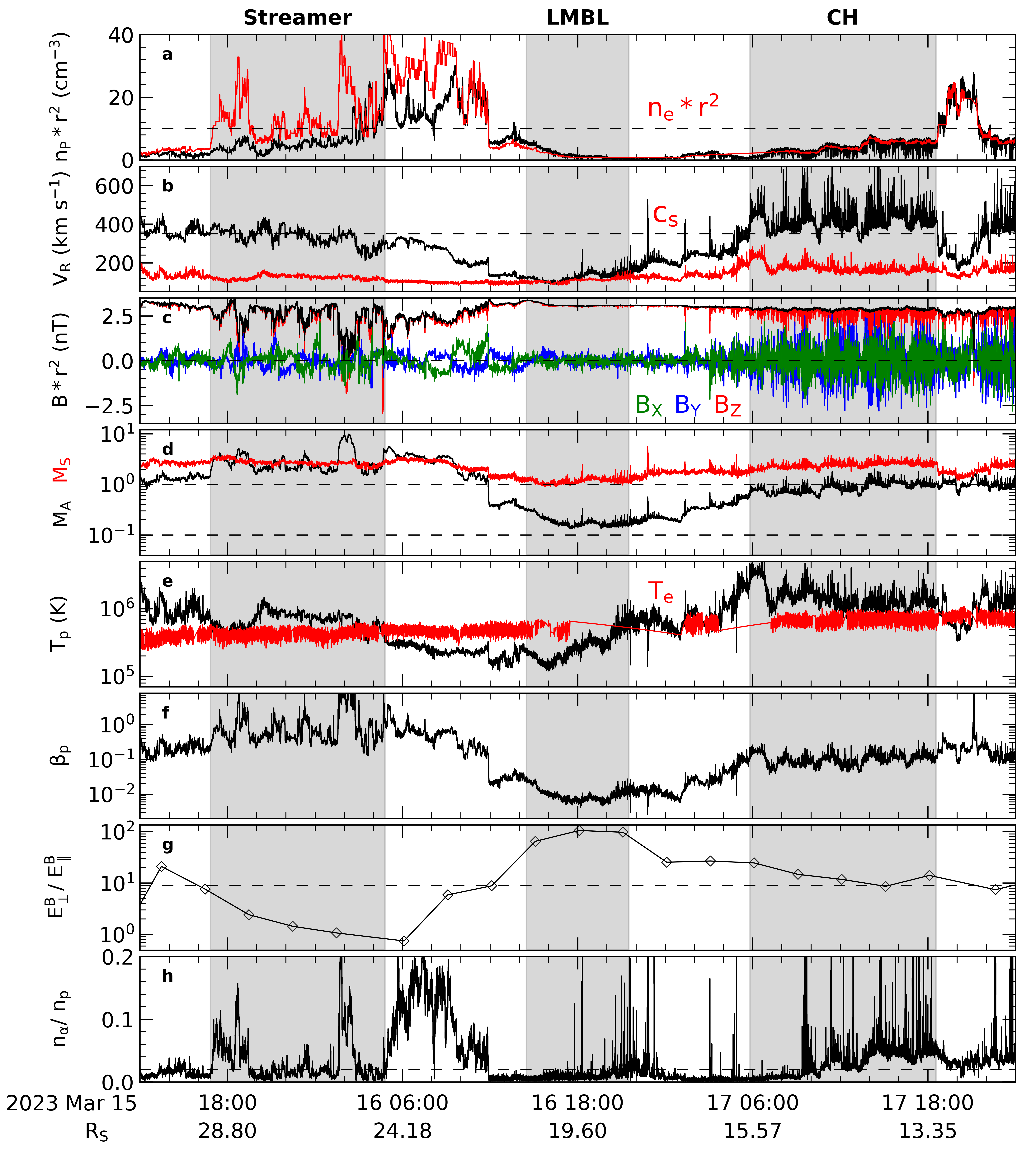}
    \caption{PSP measurements at encounter 15 from 2023 March 15 to 17 as an example of solar wind classification. The shaded areas indicate the streamer wind, the CH wind and the LMBL wind from left to right. (a) Proton density from SPAN-I and electron density from QTN (normalized to 1 au values). The horizontal dashed line marks the value of 10. (b) Proton radial velocity and the sound speed. The horizontal dashed line marks the value of 350 $\rm km\ s^{-1}$. (c) Normalized magnetic field strength and components. (d) Radial Alfvén Mach number and radial sonic Mach number. (e) Proton temperature and electron core temperature. (f) Proton plasma ${\beta_p}$. (g) Variance anisotropy of the magnetic-fluctuation spectra in the inertial range. The horizontal dashed line marks the well-known value of 9 found by \cite{1971JGR....76.3534B}. (h) Alpha-to-proton density ratio. The horizontal dashed line marks the value of 0.02. Reproduced from \cite{2024ApJ...967...58C}.}
    \label{fig:overview}
\end{figure}

As an example of the classification of the young solar wind into three source regions, Figure \ref{fig:overview} shows the measurements at encounter 15 from 2023 March 15 to 17. All three types of wind are indicated by the shaded areas. The CH wind interval has a higher velocity and a lower density, whereas the streamer wind interval has a lower velocity and a higher density (Figure \ref{fig:overview}(a) and (b)). Both the proton temperature and the electron core temperature are higher in the CH wind interval than the streamer wind interval (Figure \ref{fig:overview}(e)). Moreover, there is a difference between the CH wind interval and the streamer wind interval in the alpha-to-proton density ratio (Figure \ref{fig:overview}(h)). As for the LMBL wind, we select the near-subsonic interval reported by \cite{2024ApJ...967...58C} as the representative interval. The LMBL interval is located in the transition region from the streamer wind to the CH wind, which is consistent with its nature as boundary layers suggested by \cite{2023ApJ...944..116L}. A low density, an extremely low velocity, and a low proton temperature are obvious inside the LMBL interval as shown in Figure \ref{fig:overview}(a), (b), and (e). In addition, there is a drop in both the radial sonic Mach number and the radial Alfvén Mach number mainly caused by the extremely low velocity (Figure \ref{fig:overview}(d)). The radial sonic Mach number is defined as $M_S = {v_R} \slash {c_S}$, where ${v_R}$ is the radial solar wind velocity and ${c_S}$ is the sound speed. Similarly, the radial Alfvén Mach number is defined as $M_A = {v_R} \slash {v_A}$, where ${v_A}$ is the local Alfvén speed. The detailed calculation of ${c_S}$ and ${v_A}$ can be found in \cite{2024ApJ...967...58C}. Note that the low radial Alfvén Mach number close to 0.1 makes it inappropriate to apply Taylor’s hypothesis in the LMBL interval.

After projection to the mean field coordinate system defined in Section \ref{sec:dataset}, the magnetic field are normalized to the values at 1 au to eliminate the effects of the radial variations (Figure \ref{fig:overview}(c)). The fluctuations of $B_Z$ represent the compressible component, and fluctuations in $B_X$ and $B_Y$ represent the incompressible transverse components. The CH wind interval exhibits the overall highest fluctuation level, and the transverse fluctuations are clearly dominant compared with the compressible fluctuations. This is consistent with the observations reported by \cite{1971JGR....76.3534B} for high-speed streams. The continuous changes in the magnetic field magnitude in the streamer wind interval are likely related to a partial heliospheric current sheet (HCS) crossing, where the transverse magnetic fluctuations are less dominant. The overall lowest fluctuation level is found inside the LMBL interval. Specifically, the mean-field component of the magnetic field is extremely smooth, and compressible fluctuations essentially disappear. This may explain that the LMBL wind has the strongest variance anisotropy among all three wind types as shown in Figure \ref{fig:overview}(g). Also, the variance anisotropy of the CH wind interval is stronger than that of the streamer wind interval. The variance anisotropy $E^B_{\perp}/E^B_{\parallel}$ is given by $(P_{xx} + P_{yy})/P_{zz}$. At the same time, the $\beta_p$ is also lower in the CH wind interval and the LMBL wind interval than the streamer wind interval shown in Figure \ref{fig:overview}(f), primarily associated with a lower density. This may indicate a dependence of the variance anisotropy on the $\beta_p$ in the young solar wind, which will be further discussed in Section \ref{VA}. 

\subsection{Inertial-Range Wavevector Anisotropy}
\begin{figure}[ht]
    \centering
    \includegraphics[scale=0.75]{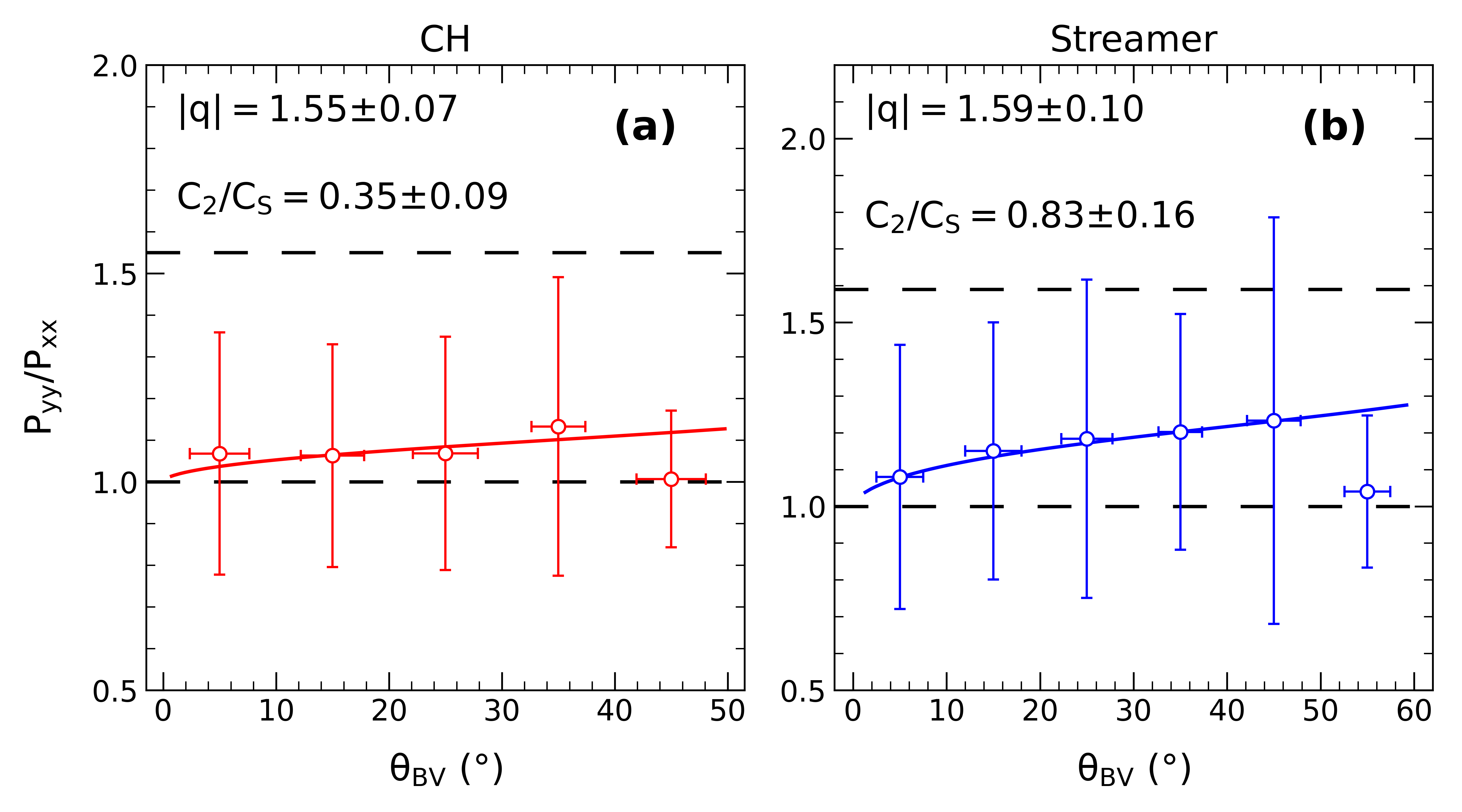}
    \caption{(a) Statistics of the ratio ${P_{yy}/P_{xx}}$ as a function of the sampling angle ${\theta_{BV}}$ for the CH wind. (b) Same for the streamer wind. The values of ${P_{yy}/P_{xx}}$ are binned every 10$\rm ^{\circ}$, and the data points and error bars are the mean values and standard deviations within the bins. The dashed lines represent the thresholds expected from the 2D and slab turbulence model. The solid curves are the model results. The average power-law index $q$ used in the model and the best-fit parameter $C_2/C_s$ for both types of the young solar wind are also listed in the figure.}
    \label{fig:super}
\end{figure}

Here we employ the \cite{1996JGR...101.2511B} analysis described in Section \ref{sec:evaluation} to examine the wavevector anisotropy in the CH wind and the streamer wind at the 1st to 19th encounters. Figure \ref{fig:super} shows the ratio $P_{yy}/P_{xx}$ in the inertial-range measurements as a function of the sampling angle $\theta_{BV}$. For both types of wind, the data are binned every 10$\rm^{\circ}$ in $\theta_{BV}$. The mean values and standard deviations of $\theta_{BV}$ and $P_{yy}/P_{xx}$ are calculated within each bin, and these values are shown as data points and error bars. In Figure \ref{fig:super}(a) and (b), the mean values of the ratio $P_{yy}/P_{xx}$ increase with the sampling angle $\theta_{BV}$ for both the CH wind and the streamer wind, except the last bin including relatively few intervals. The increasing trend is expected in the 2D and slab turbulence model. This is because only 2D component can contribute to a difference between $P_{yy}$ and $P_{xx}$ in this model, and PSP can sample more 2D fluctuations as the sampling angle increases. The range of variations of the ratio $P_{yy}/P_{xx}$ shown by error bars is larger in the streamer wind than the CH wind. This suggests that the streamer wind, which originates from closed-field regions, exhibits more variable turbulence characteristics.

For each wind type, we fit the overall dataset using Equation (\ref{bieber}) and the average power-law index $q$ to obtain a least-squares fit of $C_2/C_s$. The values of $q$ for the CH and streamer wind are given on the top of Figure \ref{fig:super}(a) and (b), respectively. The streamer wind has a slightly steeper scaling ($1.59\pm0.10$) in the inertial range than the CH wind ($1.55\pm0.07$). According to Equation (\ref{bieber}), it is clear that the ratio $P_{yy}/P_{xx}$ is expected to be 1 for pure slab turbulence (i.e., $C_2 = 0$). In contrast, for pure 2D turbulence ($C_s = 0$), the ratio $P_{yy}/P_{xx}$ should equal to the average power-law index $q$ measured in the CH wind or the streamer wind. As can be seen in Figure \ref{fig:super}, horizontal dashed lines represent the values of 1 and the average power-law index for each wind type. The mean values of the ratio $P_{yy}/P_{xx}$ are well constrained between the thresholds provided by the pure slab and pure 2D components. This once more supports that the 2D and slab turbulence model can be used to describe the wavevector anisotropy in the young solar wind observed by PSP. The best-fit model results are shown in both panels as solid curves, which agree with the measured data points reasonably well. We find $C_2/C_s = 0.35$ (or $26\%:74\%$) for the CH wind and $C_2/C_s = 0.83$ (or $45\%:55\%$) for the streamer wind. Our results indicate that the power fraction of 2D fluctuations in the inertial range for both types of wind is smaller than that reported at 1 au with $\sim 80\%$ 2D contribution \citep{1996JGR...101.2511B}. We have also tested our results, using different power-law indices in the \cite{1996JGR...101.2511B} analysis (-3/2 for the slab component and -5/3 for the 2D component). The conclusion that the slab component is dominant in the CH and streamer wind still holds. Specifically, the relative dominance of 2D component over slab component seems to be inverted in the CH wind observed by PSP. In addition, the CH wind has a larger population of slab fluctuations compared with the streamer wind. 

The dominance of slab component in the inertial range for the young solar wind within 0.3 au may provide new constraints on the initial conditions of the turbulence in the inner heliosphere, where the solar wind has less time to develop. The increase of the slab fraction in PSP measurements compared with 1 au can be explained by the anisotropic transfer of energy in the MHD turbulent cascade. In particular, the turbulent cascade mainly transfers energy to higher perpendicular wavenumbers, and eventually most energy would lie in the perpendicular wavevectors \citep[e.g.,][]{1983JPlPh..29..525S, 1995ApJ...438..763G}. This is compatible with our results that there is a greater abundance of slab component in the young solar wind observed by PSP, and the relative dominance of slab and 2D fluctuations tends to be gradually inverted via the anisotropic cascade at the inertial scales on the way to 1 au. In addition, the difference in the wavevector anisotropy between the CH wind and the streamer wind may reflect the original features of the anisotropic fluctuations established in the coronal sources. However, we still cannot conclude that this difference is exclusively determined by the source region itself, even though the solar wind fluctuations observed by PSP are the closest to the coronal sources. We also note that the difference may be caused by the dynamical evolution of the turbulence. As required by the classification criteria, the CH wind typically has a higher velocity than the streamer wind. It takes longer for the streamer wind to reach the same distances as the CH wind. As a result, the turbulence in the streamer wind has more time to evolve towards an energy distribution concentrated in the perpendicular wavevectors driven by the anisotropic cascade.

\begin{figure}[ht]
    \centering
    \includegraphics[scale=0.54]{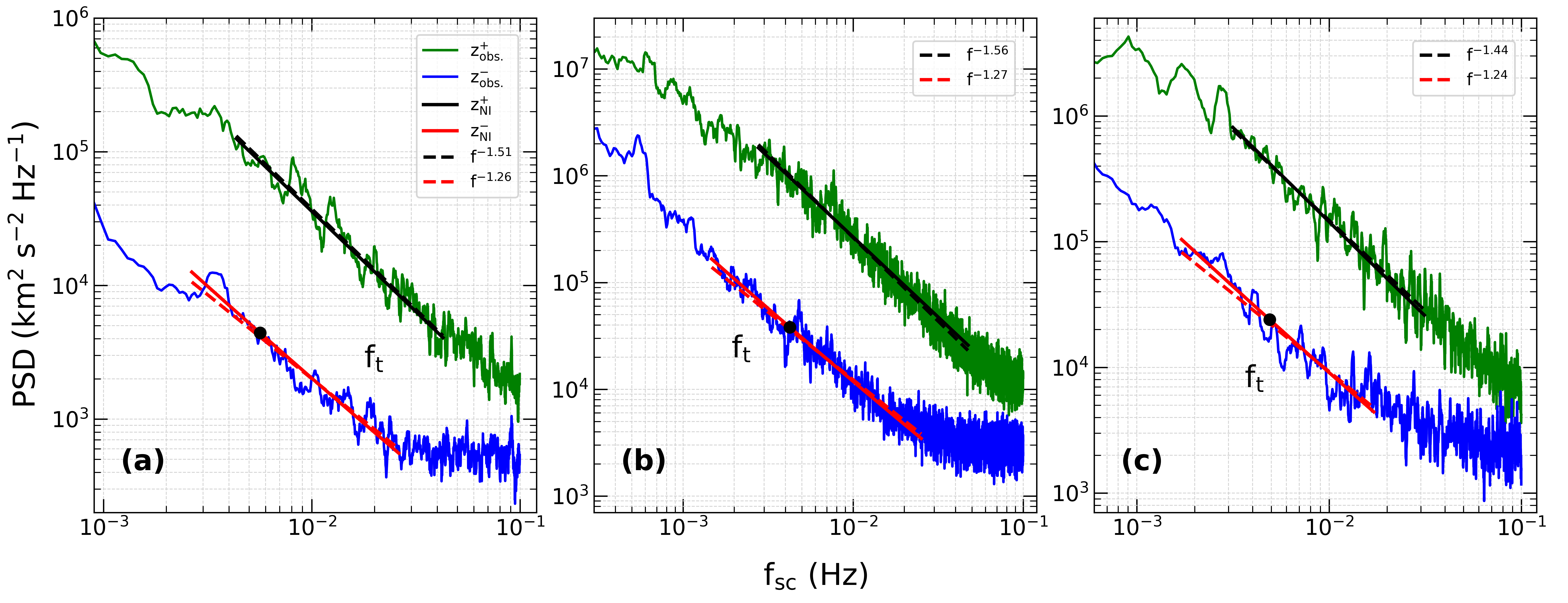}
    \caption{PSDs of the Elsässer variables $\textit{\textbf{z}}^{\pm}$ for the near-subsonic intervals. (a) Encounter 10. (b) Encounter 13. (c) Encounter 15. The solid black and red lines in each panel show the theoretical spectra predicted by the NI MHD turbulence model. The dashed black and red lines indicate the power-law fits to the observed spectra. $f_{t}$ denotes the transition frequency (see text for details).}
    \label{fig:wavenumber}
\end{figure}

We further investigate the wavevector anisotropy in the near-subsonic intervals and oblique sub-Alfvénic intervals, which serve as a representation of the LMBL wind. Figure \ref{fig:wavenumber} gives the trace PSDs of the forward and backward Elsässer variables $\textit{\textbf{z}}^{\pm}$ for the near-subsonic intervals at encounter 10, 13, and 15. As shown in Figure \ref{fig:wavenumber}(a), the outward-propagating $\textit{\textbf{z}}^{+}$ fluctuations dominate during the near-subsonic interval at encounter 10, with the spectral amplitude nearly one order of magnitude larger than that of the $\textit{\textbf{z}}^{-}$ mode. In addition, the $\textit{\textbf{z}}^{+}$ spectrum follows a single power law in frequency throughout the inertial range, while the $\textit{\textbf{z}}^{-}$ spectrum gradually flattens at higher frequencies exhibiting a concave shape. Similar behaviors of the $\textit{\textbf{z}}^{\pm}$ spectra can be seen in the near-subsonic intervals at encounter 13 and 15 (Figure \ref{fig:wavenumber}(b) and (c)). The flattening of the $\textit{\textbf{z}}^{-}$ spectra compared with those of $\textit{\textbf{z}}^{+}$ has been reported in a previous sub-Alfvénic interval observed by PSP \citep[e.g.,][]{2022ApJ...926L..16Z}. The different behaviors of the $\textit{\textbf{z}}^{+}$ and $\textit{\textbf{z}}^{-}$ spectra may suggest that the contribution of 2D fluctuations is not significant to the observed spectrum, which would otherwise result in the same spectral shape for both modes in the NI MHD turbulence model. 

To accurately evaluate the contribution of the slab and 2D components, we construct the solid black and red lines in Figure \ref{fig:wavenumber} as theoretical predictions of the $\textit{\textbf{z}}^{+}$ and $\textit{\textbf{z}}^{-}$ spectra using Equations (\ref{k+}) and (\ref{k-}), respectively. We adopt the same fitting range in the wavenumber space for both $\textit{\textbf{z}}^{\pm}$ spectra, satisfying the following criteria: a broad wavenumber range enables the $\textit{\textbf{z}}^{-}$ spectrum to exhibit an entire concave shape, with its end set at the points where the $\textit{\textbf{z}}^{-}$ spectrum starts to flatten possibly due to the instrumental noise. As shown in Figure \ref{fig:wavenumber}, power-law fits in the same wavenumber range give the power-law indices of the $\textit{\textbf{z}}^{\pm}$ spectra, which agree with the finding of \cite{2024ApJ...966..144W}. In addition, the spectra of $\textit{\textbf{z}}^{\pm}$ predicted by the NI MHD theory nearly coincide with the power-law fits to the observed spectra. This indicates that the NI MHD turbulence model well interprets the observations of the near-subsonic intervals. The corresponding fit parameters are listed in Table \ref{tab:table}. As for all the near-subsonic intervals, both spectra of $\textit{\textbf{z}}^{\pm}$ are dominated by slab fluctuations in this model, as indicated by the ratio $C^{*\pm} / C^{\infty} \gg 1$. Furthermore, the amplitude of the forward slab fluctuations is at least one order larger than that of the backward modes. This may suggest that all the near-subsonic intervals are populated by unidirectionally propagating Alfvén waves, which can induce energy cascade of imbalanced Alfvénic turbulence as demonstrated by \cite{2023NatCo..14.7955Y}. A similar conclusion is reached by \cite{2024ApJ...972..129E}. 

Note that the dominance of slab fluctuations in the near-subsonic interval at encounter 15 is likely due to the nearly parallel sampling angle $\theta_{BV}$ ($\sim 161 \rm ^{\circ}$). To ensure that both the slab and 2D components are important in the observed turbulence, we use the criterion $ 30\rm ^{\circ} < $ $\theta_{BV}$ $ < 60\rm ^{\circ}$ ($ 120\rm ^{\circ} < $ $\theta_{BV}$ $ < 150\rm ^{\circ}$) to identify the oblique sub-Alfvénic intervals as inspired by \cite{2022ApJ...933...56A}. We repeat the same procedure of applying the NI MHD spectral theory to the $\textit{\textbf{z}}^{\pm}$ spectra in the oblique sub-Alfvénic intervals, and the model parameters are also summarized in Table \ref{tab:table}. Again, the oblique sub-Alfvénic intervals, akin to the near-subsonic intervals, show the dominance of slab fluctuations for both $\textit{\textbf{z}}^{\pm}$ spectra with the ratio $C^{*\pm} / C^{\infty} \gg 1$. The slab component is again mainly composed of the forward slab fluctuations, as indicated by the ratio $C^{*+} / C^{*-} \geq 8$. Hence, it is worth noting that the spectral shape of the $\textit{\textbf{z}}^{\pm}$ modes is primarily reproduced by the slab component in the NI MHD turbulence model. Such similar wavevector anisotropy in these intervals may indicate that the same basic turbulence physics holds in both the near-subsonic intervals and oblique sub-Alfvénic intervals, given their common source region from LMBLs.

\subsection{Inertial-Range Variance Anisotropy} \label{VA}
\begin{figure}[ht]
    \centering
    \includegraphics[scale=0.93]{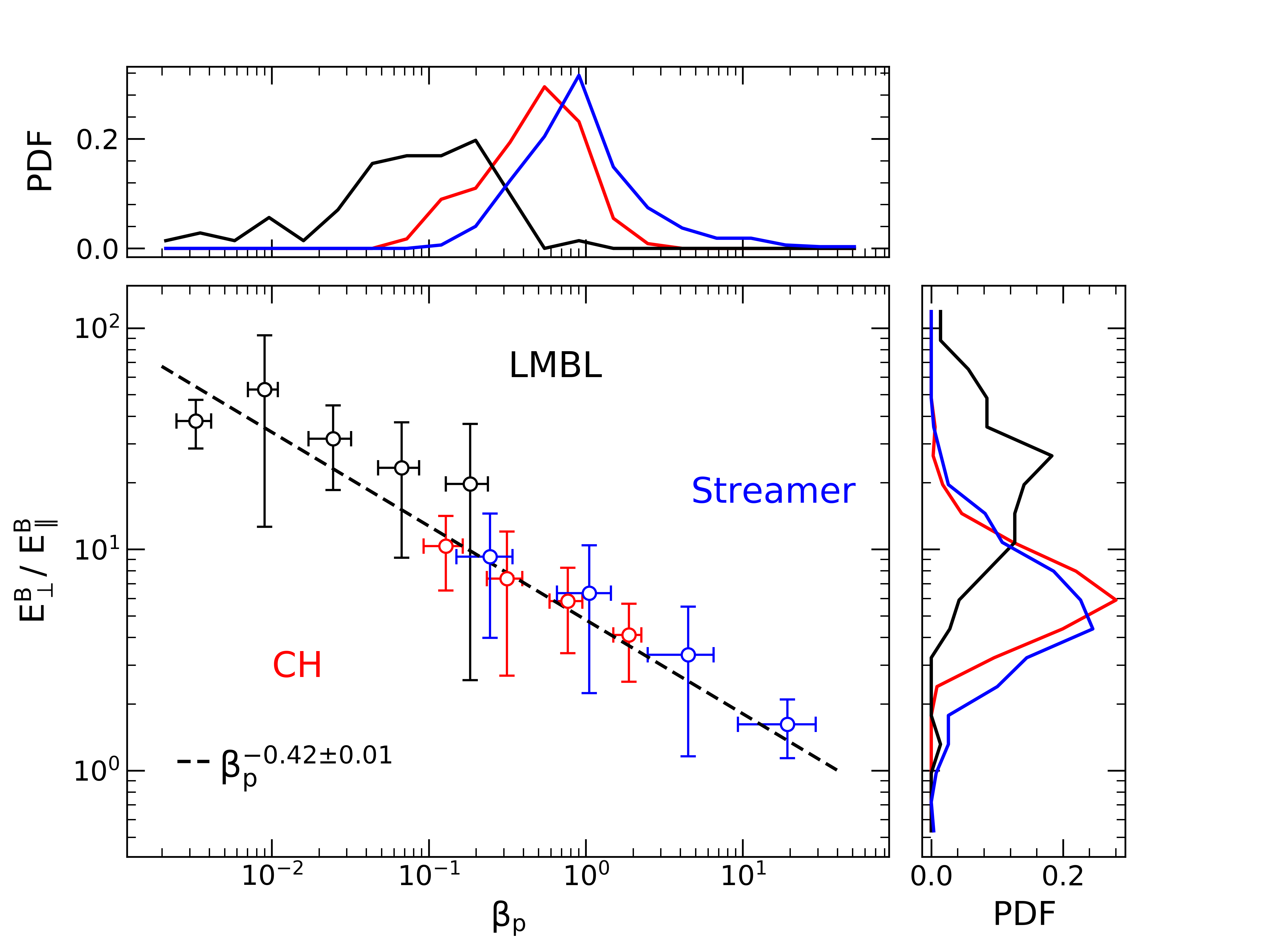}
    \caption{Variance anisotropy $E_{\perp}^{B}/E_{\parallel}^{B}$ in the inertial range as a function of $\beta_p$. The values of $E_{\perp}^{B}/E_{\parallel}^{B}$ are binned by $\beta_p$ in the logarithmic space. Red is for the CH wind, black is for the LMBL wind, and blue is for the streamer wind. The dashed black line denotes the power-law fit of the relation with $\beta_p$ for all the young solar wind, and the power-law exponent is listed. The top and right panels give the PDFs of $\beta_p$ and $E_{\perp}^{B}/E_{\parallel}^{B}$.}
    \label{fig:beta}
\end{figure}

Using the solar wind data selected from different source regions, we show in Figure \ref{fig:beta} the variance anisotropy $E_{\perp}^{B}/E_{\parallel}^{B}$ in the inertial range as a function of $\beta_p$. For each wind type, we bin the data equally within the respective logarithmic range of $\beta_p$, and the mean values and standard deviations are shown as the data points and error bars. In the top panel, the overall distribution of $\beta_p$ shifts towards lower values in the CH wind and the LMBL wind compared with the streamer wind. This is mainly attributed to the low density limited in the classification criteria for the CH wind and the LMBL wind. The highest $\beta_p$ in the streamer wind arises from its higher density and lower magnetic field magnitude, which is associated with the partial HCS crossing as shown in Figure \ref{fig:overview}. In addition, the streamer wind, characterized by the highest $\beta_p$, exhibits the lowest variance anisotropy in the inertial range (main panel of Figure \ref{fig:beta}). This indicates a greater percentage of compressible fluctuations in the streamer wind. As shown in the right panel, the LMBL wind has the largest overall variance anisotropy among all three wind types, consistent with the results of \cite{2022ApJ...926L...1B}. The overall variance anisotropy of the CH wind is slightly lager than that of the streamer wind. The probability density function (PDF) for the CH wind peaks at around $6$, which is lower than the well-known anisotropy of $9$ found in the high‐speed streams at 1 au \citep{1971JGR....76.3534B}. 

Consistent with the finding of \cite{2006JGRA..111.9111S}, we find a dependence of the variance anisotropy in the inertial range on the $\beta_p$ for the observed young solar wind inside 0.3 au. For three different wind types, this dependence can be described by a common power-law function as shown in the main panel of Figure \ref{fig:beta}, which follows a relation of $\sim \beta_p^{-0.42\pm 0.01}$. This may indicate that near-Sun turbulence tends to be more compressive as $\beta_p$ increases. Likewise, we perform the fitting for each type of wind (not shown here). Similar power-law exponents are obtained with $-0.34\pm 0.07$ for the LMBL wind, $-0.37\pm 0.03$ for the CH wind, and $-0.41\pm 0.03$ for the streamer wind. The inverse relationship between the $\beta_p$ and the variance anisotropy is also consistent with the simulations of \cite{1996JGR...101.7619M}, but it remains to be understood theoretically. We note that the variance anisotropy of the CH wind is nearly in agreement with that of the streamer wind of similar $\beta_p$. This suggests that in situ dynamics, rather than the solar origin, may have an effect on the variance anisotropy in the young solar wind, as suggested by \cite{2006JGRA..111.9111S}. However, the $\beta_p$ at extreme values can still be used as an indicator to the source regions of the young solar wind, since different solar wind streams have not yet interacted significantly at PSP encounter distances. At extreme values of $\beta_p$, the LMBL wind and the streamer wind exhibit the strongest and weakest variance anisotropy, respectively. This may suggest that the variance anisotropy can be viewed as the original features of fluctuations shaped by coronal sources for the young solar wind with extreme $\beta_p$.

\section{Conclusions} \label{sec:Conclusion}
In this Letter, using the first 19 encounters of PSP measurements, we have identified the young solar wind from different source regions: coronal hole interiors, streamers, and the inner boundaries of coronal holes (i.e., LMBLs). Key findings are revealed concerning the wavevector and variance anisotropies of each type of wind. We summarize the main results as follows.
\begin{enumerate}
    \item Based on the 2D and slab turbulence model, we find that the power fraction of 2D fluctuations in the inertial range for the CH wind and the streamer wind is smaller than the $\sim80\%$ reported at 1 au. Specifically, the relative abundance between slab and 2D fluctuations seems to be inverted for the CH wind observed by PSP. A probable explanation for the increase of the slab fraction is the anisotropic cascade in the inertial range on the way to 1 au. In addition, the CH wind has a larger population of slab fluctuations than the streamer wind. The difference between them may reflect the original features of the fluctuations from their coronal sources.
    \item In the context of the NI MHD turbulence model, we investigate the wavevector anisotropy in the near-subsonic intervals and oblique sub-Alfvénic intervals, which are representative of the LMBL wind. The modeling results show that the oblique sub-Alfvénic intervals, akin to the near-subsonic intervals, exhibit the dominance of slab fluctuations for both $\textit{\textbf{z}}^{\pm}$ spectra. Also, the forward slab fluctuations dominate, with the spectral amplitude at least $8$ times larger than that of the backward modes. This indicates that the near-subsonic intervals and the oblique sub-Alfvénic intervals are populated by unidirectionally propagating Alfvén waves. All the results suggest a greater fraction of slab fluctuations in the young solar wind compared with 1 au observations.
    \item We find that the variance anisotropy in the inertial range scales with $\beta_p$ in the young solar wind, following a power-law dependence $\sim \beta_p^{-0.42\pm 0.01}$. This relation likely indicates that magnetic compressibility tends to be higher as $\beta_p$ increases. Moreover, the largest overall variance anisotropy is observed in the LMBL wind. The overall variance anisotropy of the CH wind is slightly lager in comparison with that of the streamer wind. Note that the LMBL wind and the streamer wind exhibit the strongest and weakest variance anisotropy at extreme $\beta_p$, respectively. This contrast can be viewed as the remnant of turbulence properties from their coronal source regions.
\end{enumerate}

%\begin{acknowledgments}

The research was supported by the Strategic Priority Research Program of the Chinese Academy of Sciences (No. XDB 0560000), NSFC (under grants 42474223, 42274201, 12073032, 42150105, and 42274213), the National Key R\&D Program of China (No. 2022YFF0503800 and No. 2021YFA0718600), and the Specialized Research Fund for State Key Laboratories of China. H.R. is supported by a studentship from the UK Science and Technology Facilities Council (STFC). We acknowledge the NASA Parker Solar Probe mission and the SWEAP and FIELDS teams for the use of data.

%\end{acknowledgments}

\bibliography{sample631}{}
\bibliographystyle{aasjournal}

 \newpage

\newpage

\newpage

\newpage

\begin{deluxetable*}{ccccccccccccc}

\tablecaption{Parameters Related to the NI MHD Turbulence Model.}
\label{tab:table}
\tablehead{
\colhead{No.} & \colhead{Enc.} & \colhead{Start} & \colhead{Duration} & \colhead{$\theta_{BV}$} & \colhead{$C^{*}+$} & \colhead{$C^{*}-$} & \colhead{$C^{\infty}$} & \colhead{$f_{t}$}
\\
\colhead{} & \colhead{} & \colhead{(UT)} & \colhead{(hr)} &\colhead{($\rm ^{\circ}$)} & \colhead{} & \colhead{} & \colhead{}  & \colhead{(Hz)} 
\\
\colhead{(1)} & \colhead{(2)} & \colhead{(3)} & \colhead{(4)} &\colhead{(5)} & \colhead{(6)} & \colhead{(7)} & \colhead{(8)} & \colhead{(9)}}

\startdata
1 & $10^{*}$ & 2021-11-21 21:24 & 3.3 & 119.8 & 4.93 & 0.23 & $4.52 \times 10^{-30}$ & $5.64 \times 10^{-3}$ \\
2 & $13^{*}$ & 2022-09-06 19:50 & 7.1 & 151.5 & 23.28 & 0.89 & $7.72 \times 10^{-25}$ & $4.25 \times 10^{-3}$ \\
3 & $15^{*}$ & 2023-03-16 14:34 & 6.9 & 161.2 & 11.78 & 0.65 & $6.28 \times 10^{-23}$ & $4.91 \times 10^{-3}$ \\
4 & 10 & 2021-11-22 03:20 & 3 & 126.9 & 6.54 & 0.03 & $1.60 \times 10^{-25}$ & $4.58 \times 10^{-7}$ \\
5 & 10 & 2021-11-22 06:20 & 3 & 138.6 & 11.69 & 0.08 & $9.59 \times 10^{-34}$ & $6.96 \times 10^{-5}$ \\
6 & 12 & 2022-06-02 20:18 & 3 & 133.4 & 3.18 & 0.34 & $8.72 \times 10^{-17}$ & $4.92 \times 10^{-3}$ \\
7 & 12 & 2022-06-02 23:18 & 3 & 142.1 & 9.32 & 0.06 & $9.22 \times 10^{-26}$ & $1.75 \times 10^{-7}$ \\
8 & 12 & 2022-06-03 02:18 & 3 & 145.9 & 12.05 & 0.46 & $1.14 \times 10^{-21}$ & $3.24 \times 10^{-3}$ \\
9 & 13 & 2022-09-06 09:30 & 3 & 35.7 & 65.43 & 0.21 & $1.66 \times 10^{-17}$ & $8.59 \times 10^{-7}$ \\
10 & 13 & 2022-09-07 02:52 & 3 & 129.7 & 4.63 & 0.52 & $1.04 \times 10^{-20}$ & $2.62 \times 10^{-2}$ \\
\enddata
\tablecomments{Columns (1)-(5) correspond to the number, encounter number, start time, duration, and the angle between the mean magnetic field and mean velocity in the spacecraft frame of the near-subsonic intervals and the oblique sub-Alfvénic intervals, respectively. Columns (6)-(9) give the power of forward-propagating slab fluctuations, the power of backward-propagating slab fluctuations, the power of 2D fluctuations, and the transition frequency. The * superscript marks the near-subsonic intervals. }
\end{deluxetable*}

\end{document}